# ABOUT THE ELECTRODYNAMIC ACCELERATION OF MACROSCOPIC PARTICLES

*S. N. Dolya, K.A. Reshetnikova*

*Joint Institute for Nuclear Research, Joliot – Curie Street 6, Dubna, Russia, 141980*
*E – mail: sndolya@jinr.ru*

**Abstract**

An electric charge is imparted to macroscopic particles, whereupon they are pre-accelerated in an electrostatic field by the high voltage U = 220 kV. Then the particles are accelerated by a traveling electromagnetic wave with the initial phase velocity lying in the range $\beta = 10^{-3} – 10^{-5}$. Focusing the particles is provided by electrostatic doublets. At the acceleration length L = 20 m, the particles with $Z/A = 2.3*10^{-7}$ increase their velocity from $\beta = 10^{-5}$ to $\beta = 10^{-4}$.

## 1. Introduction

Below, acceleration of very small balls with the diameter D = 0.1-100 microns by the electrodynamic method is studied. These accelerated small balls can be used for imitation of damage to cosmic objects by micrometeorites [1].

In order to accelerate these neutral small balls, an electric charge must be placed on them. Then the charged balls can be accelerated by the electromagnetic wave. Consideration is given to acceleration of carbon small balls with the density $\rho = 2.3$ g/cm$^3$. Using simple calculations, one can see that the density of nucleons in carbon is $N_n = 1.38*10^{24}$ n/cm$^3$.

## 2. The electric charge on the small balls

The electric field on the surface of a small ball can be found from the Coulomb equation: $E_{surf} = Ze/r^2$. Let us equate this field to the threshold field $E_{th}$ and then estimate what maximal charge can be placed on a small ball. For the auto electron emission [1] the threshold field is $E^e_{th} = 10^9$ V/cm, for the auto ion emission $E^i_{th} = 10^{10}$ V/cm. Assuming $E_{surf} = E_{th}$, one can find the maximal charge on a small ball:

$$Ze/r^2 = E^i_{th} . \qquad (1)$$

From (1), the number N, where Z = Ne (e is an elementary charge), can be calculated. From equation $M_b = AM_n$, where $M_b$ is the mass of a small ball, $M_n$ is the nucleon mass, let us estimate A, which is the number of nucleons in a small ball. Now, one can find the charge-to-mass ratio for small balls Z/A. For small balls this ratio lies within the range $Z/A = 2.3*10^{-3} - 2.3*10^{-7}$; for a single-charge uranium ion ($U^{+1}_{238}$) this ratio is $Z/A = 1/238 = 4.2*10^{-3}$.

After pre-acceleration in an electrostatic field with the high voltage $U_{el.st} = 220$ kV, these small balls will have the velocity $\beta$ lying in the range $\beta = 10^{-3} - 10^{-5}$, where velocity is measured in units of light velocity in vacuum $\beta = V/c$, $c = 3*10^{10}$ cm/s. The parameters of the accelerated small balls are given in Table 1.

Table 1. The parameters of the accelerated small balls

| Diameter | A | Z | Mass, g | Z/A | $\beta$, $U_{el.st} = 220$ kV |
|---|---|---|---|---|---|
| 0.1 μ | $7.2*10^8$ | $1.66*10^6$ | $1.2*10^{-15}$ | $2.3*10^{-3}$ | $10^{-3}$ |
| 10 μ | $7.2*10^{14}$ | $1.66*10^{10}$ | $1.2*10^{-9}$ | $2.3*10^{-5}$ | $10^{-4}$ |
| 1 mm | $7.2*10^{20}$ | $1.66*10^{14}$ | $1.2*10^{-3}$ | $2.3*10^{-7}$ | $10^{-5}$ |
| 100 mm | $7.2*10^{26}$ | $1.66*10^{18}$ | $1.2*10^3$ | $2.3*10^{-9}$ | $10^{-6}$ |

## 3. Requirements to slowing-down of the electromagnetic wave

In order to accelerate a particle in a traveling wave, the initial velocities of the particle and wave must be equal $\beta = \beta_{ph}$ and the phase velocity of the wave during acceleration must be increased.

A slow electromagnetic wave with the phase velocity $\beta_{ph}^{-1} = 10^2 - 5*10^2$ can be achieved in a spiral waveguide. As is known [2], slowing-down of the electromagnetic wave in a spiral waveguide is simply the ratio of the helix perimeter to the helix pitch distance. The perimeter of the helix is $2\pi r_0$, the helix pitch distance is h and the dispersion equation is

$$\beta_{ph} = tg\Psi, \qquad (2)$$

where $\beta_{ph} = V_{ph}/c$ and $tg\Psi = h/2\pi r_0$ are the tangents of the angle of the spiral waveguide winding. If the radius of the helix is chosen to be $r_0 = 10$ cm and the helix pitch distance $h = 0.1$ cm, one obtains $\beta_{ph} = 1/628 = 0.0016$. This expression shows the well-known independence of the phase velocity from the frequency.

It is also known that the medium with a big value of the dielectric $\varepsilon$ and magnetic $\mu$ permeability slows down the electromagnetic wave in the following way:

$$\beta_{ph} = 1/(\varepsilon\mu)^{1/2}, \qquad (3)$$

so, one can expect that when the spiral waveguide is placed in such a medium, the phase velocity will be:

$$\beta_{ph} = tg\Psi /(\varepsilon\mu)^{1/2}, \qquad (4)$$

where tg$\Psi$ determines slowing-down of the spiral waveguide and $(\varepsilon\mu)^{1/2}$ shows slowing-down qualities of the medium. Thus, for the big values $(\varepsilon\mu)^{1/2}$ and small value tg$\Psi$, according to formula (4), very huge slowing-down can be achieved.

**4. Accurate decisions**

Following [2], the properties of the spiral waveguide fully immersed in a medium with the dielectric permeability $\varepsilon$ and magnetic permeability $\mu$ will be considered. Let us find the components of the electric and magnetic fields for the inside area of the spiral:

$$E_{z1} = E_0 I_0(k_1 r)$$
$$E_{r1} = i(k_3/k_1) E_0 I_1(k_1 r)$$
$$H_{\varphi 1} = i\varepsilon(k/k_1) E_0 I_1(k_1 r)$$
$$H_{z1} = -i(k_1/\mu k)\text{tg}\Psi I_0(k_1 r_0) E_0 I_0(k_1 r)/I_1(k_1 r_0) \quad (5)$$
$$E_{\varphi 1} = -\text{tg}\Psi I_0(k_1 r_0) E_0 I_1(k_1 r)/I_1(k_1 r_0)$$
$$H_{r1} = (k_3/\mu k)\text{tg}\Psi I_0(k_1 r_0) E_0 I_1(k_1 r)/I_1(k_1 r_0),$$

and the components of the electric and magnetic fields for the outside area:

$$E_{z2} = I_0(k_1 r_0) E_0 K_0(k_1 r)/K_0(k_1 r_0)$$
$$E_{r2} = -i(k_3/k_1) I_0(k_1 r_0) E_0 K_1(k_1 r)/K_0(k_1 r_0)$$
$$H_{\varphi 2} = -i(k/k_1) I_0(k_1 r_0) E_0 K_1(k_1 r)/K_0(k_1 r_0)$$
$$H_{z2} = i(k_1/k)\text{tg}\Psi I_0(k_1 r_0) E_0 K_0(k_1 r)/K_1(k_1 r_0) \quad (6)$$
$$E_{\varphi 2} = -\text{tg}\Psi I_0(k_1 r_0) E_0 K_1(k_1 r)/K_1(k_1 r_0)$$
$$H_{r2} = (k_3/\mu k)\text{tg}\Psi I_0(k_1 r_0) E_0 K_1(k_1 r)/K_1(k_1 r_0),$$

where, as in [2], $e^{i(\omega t - k_3 z)}$ is not written.

The dispersion equation relating the phase velocity and frequency for a spiral waveguide fully immersed in a ferrodielectric is:

$$\text{ctg}^2\Psi = k_1^2/k^2 \{I_0(k_1 r_0) K_0(k_2 r_0)/I_1(k_1 r_0) K_1(k_2 r_0)\}, \quad (7)$$

where $k_1 = k(1/\beta_{ph}^2 - 1)^{1/2}$, $k_2 = k(1/\beta_{ph}^2 - \varepsilon\mu)^{1/2}$, $k = \omega/c$.

For a big slowing-down, the dispersion equation simplifies to:

$$\beta_{ph} = \text{tg}\Psi/(\varepsilon\mu)^{1/2}, \quad (8)$$

which coincides with the expected equation (4).

This formula, as for a spiral waveguide in vacuum [2], has a simple physical meaning: slowing-down of the phase velocity of the electromagnetic wave is a ratio of the way the electromagnetic wave goes along the spiral to the way the electromagnetic wave travels along the spiral axis.

Using simple calculations, the relation between the power flux in a spiral waveguide and the electric field on the spiral waveguide axis is obtained for a waveguide fully immersed in a ferrodielectric medium:

$$P = (c/8) E_0^2 r_0^2 [kk_3/k_1^2] \varepsilon\{(1+I_0K_1/I_1K_0)(I_1^2-I_0I_2) +$$

$$+ (I_0/K_0)^2(1+I_1K_0/I_0K_1)(K_0K_2-K_1^2)\}, \qquad (9)$$

where $k_3 = \omega/V_{ph}$.

This relation fully coincides with the expression for a spiral waveguide in vacuum [2], except for the multiplier $\varepsilon$ determining the dielectric permeability.

## 5. Ferrodielectric medium

Let us assume that a spiral waveguide with the radius $r_0$ is fully immersed in a ferrodielectric medium. First and foremost, this medium must have a small tangent $tg\delta \ll 1$ because otherwise the electromagnetic wave might damp in the medium. Secondly, the magnetic induction $B_{work}$ must not be greater than

$$B_{work} < 0.2 \text{ T}. \qquad (10)$$

It is connected with the complex dependence of the magnetic induction B inside the ferromagnetic on the magnetic field H. A larger induction in the spiral waveguide arises at the beginning of the accelerating section at the entrance of the RF power to the section. Using formulae (5) and (6), let us find the relation between the magnetic induction at that point and the electric field on the axis of the spiral waveguide:

$$B = \mu H_z = itg\Psi E_0(k_1/k)I_0^2(k_1r_0)/I_1(k_1r_0) < B_{work}. \qquad (11)$$

Now, the relation for $E_0$ and B can be obtained:

$$E_0 < B_{work} I_1(k_1r_0)/itg\Psi(k_1/k)I_0^2(k_1r_0), \qquad E_0 < k(\beta_{ph})*B_{work}, \qquad (12)$$

where $k(\beta_{ph})$ is the coefficient quickly decreasing with the phase velocity (see permissible values of the electric power in Table 2,3).

Table 2. Parameters of the sections.

| Parameter | Section 1 | Section 2 |
|---|---|---|
| $Z/A = 2.3*10^{-3}$, $U_{el.st.} = 220$ kV, $(\varepsilon\mu)^{1/2} = 10$, $V_{initial} = 300$ km/s | $P = 110$ kW, $\varepsilon, \mu = 10$. Fully filled with a ferrodielectric | $P = 200$ kW, $\varepsilon, \mu = 10$. Fully filled with a ferrodielectric |
| Velocity, initial – final, $\beta_{ph}$ | $10^{-3}$ - $3.16*10^{-3}$ | $3.16*10^{-3}$-$10^{-2}$ |
| Starting - ending spiral radius, $r_0$ | 3 – 1.81 cm | 2.41 – 1.35 cm |
| Frequency $f_0$, Hz | $3.333*10^6$ | $12.5*10^6$ |
| Average electric field strength, $\bar{E}_0$ | 6 kV/cm | 6 kV/cm |
| Maximal magnetic induction, $B_{max} = \mu H_z$ | 0.1982 T | 0.197 T |
| Section length | 3.8 m | 43 m |

For real-world acceleration of particles, it is necessary to open a vacuum channel for particles. The formulas corresponding to this case are very lengthy and therefore not given here. If the radius of the channel is small enough, compared to the radius of the waveguide, the motion of the particles will not change significantly, compared to the waveguide fully filled with a ferrodielectric. All the values shown in the tables and figures have been obtained using accurate formulas, the approximate formulas being provided only with a view to explain the physical meaning.

The accelerator consists of two sections, each section being fed by the individual RF power and individual frequency. The spiral is wound onto a tapering cone [3,4], where the starting radius is bigger than the ending one. Due to this, the electric field strength is approximately uniform along the acceleration section. In case of a cylindrical radius waveguide, the helix pitch distance would increase drastically and the electric field strength would decrease quickly along the section.

At each stage of calculations, three equations are solved simultaneously: a dispersion equation (7), an equation for power flux that relates phase velocity with electric field strength (9), and the equation of motion for particles

$$dV/dt = (Ze/AM_n)E_0\cos\varphi_s, \qquad (13)$$

where $\cos\varphi_s = 0.7$, $\varphi_s = 45^0$ is the synchronic phase.

If the magnetic permeability is increased up to $\mu = 25.6$, a larger slowing-down can be achieved, Table 3.

Table 3. Parameters of the sections.

| Parameter | Section 1 | Section 2 |
|---|---|---|
| $Z/A = 2.3*10^{-5}$<br>$U_{el.st} = 220$ kV, $(\varepsilon\mu)^{1/2} = 16$<br>$V_{initial.} = 30$ km/s | $P = 11.8$ kW<br>$\varepsilon =10$, $\mu = 25.6$<br>Fully filled with a ferrodielectric | $P = 15$ kW<br>$\varepsilon =10$, $\mu = 25.6$<br>Fully filled with a ferrodielectric |
| Velocity range, initial – final, $\beta_{ph}$ | $10^{-4}$ -$3.16*10^{-4}$ | $3.16*10^{-4}$-$10^{-3}$ |
| Starting – ending spiral radius, $r_0$ | 10 – 5.5 cm | 5.5 – 3 cm |
| Frequency $f_0$, Hz | $10^5$ | $6*10^5$ |
| Average electric field strength, $\bar{E}_0$ | 2 kV/cm | 2 kV/cm |
| Maximal magnetic induction, $B_{max} = \mu H_z$ | 0.1732 T | 0.1961 T |
| Section length | 12 m | 120 m |

If the magnetic permeability μ is increased a hundredfold up to μ = 2560, one can achieve the slowing-down of the electromagnetic wave by the medium $(\varepsilon\mu)^{1/2} = 160$, but the average electric field strength will be very low in this case. The dependence of the magnetic permeability μ on the magnetic field strength H, μ = μ(H), makes it inconvenient to use a ferrodielectric medium for filling a spiral waveguide in our case.

## 6. Partial filling of the spiral waveguide by the medium

In the case when a ferrodielectric is placed outside the spiral waveguide and there is vacuum inside the waveguide, simple analytic formulas can be obtained. Formula (9) relating the power flux with the electric field strength on the axis of the spiral can be written as:

$$P = (c/8)\, E_0^2 r_0^2 [\, kk_3/k_1^2 \,]\{(1+I_0K_1/I_1K_0)(I_1^2-I_0I_2) + $$

$$+ \varepsilon\, (I_0/K_0)^2 (1+I_1K_0/I_0K_1)(K_0K_2-K_1^2)\}. \quad (14)$$

Only the second addend, corresponding to the power flux moving outside the spiral, multiplies by the dielectric permeability ε. Let us refer to this case as "partial filling of the spiral waveguide by the medium".

The dispersion equation for this case is:

$$\text{ctg}^2\Psi = (k_1k_2/k^2)\varepsilon\mu\{I_0(k_1r_0)K_0(k_2r_0)/I_1(k_1r_0)K_1(k_2r_0)\}F_0, \quad (15)$$

where

$$F_0 = \varepsilon\{1+(k_1/k_2)\mu\, I_0K_1/I_1K_0\}*[1+(k_1/k_2)\varepsilon\, I_0K_1/I_1K_0]^{-1}. \quad (16)$$

The argument of the functions $I_{0,1}$ is $(k_1 r_0)$ and the argument of the functions $K_{0,1}$ is $(k_2 r_0)$. The most interesting case is the large slowing-down of the electromagnetic wave, when the dispersion equation simplifies to look like formula (4) and can be written as follows:

$$\beta_{ph} = \text{tg}\Psi \, F_0^{1/2}/(\varepsilon\mu)^{1/2}. \qquad (17)$$

In the important case $\varepsilon, \mu \gg 1$, the formula for $F_0$ becomes simpler:

$$F_0 = \mu, \qquad (18)$$

and the dispersion equation transforms into

$$\beta_{ph} = \text{tg}\Psi/\varepsilon^{1/2}. \qquad (19)$$

On the one side, it is easier, in this case, to achieve the requirement to the magnetic induction inside the ferrite because the magnetic field amplitude outside the spiral is much lower than that inside the spiral. On the other side, slowing-down of the wave by the medium is worse by a factor of $\mu^{1/2}$ than the full filling of the waveguide with a ferrodielectric medium.

The most interesting case is when a non-magnetic dielectric with a big dielectric permeability $\varepsilon$, $\varepsilon \gg 1$, is placed outside the spiral and the medium is non-magnetic, $\mu = 1$, the formula for $F_0$ becomes very simple, $F_0 = 2$, and the dispersion equation can be written as:

$$\beta_\phi = \sqrt{2} \, \text{tg}\Psi/\varepsilon^{1/2}. \qquad (20)$$

This case, allowing one to receive a large slowing-down (simple formula (20)) and great electric field strength (formula (14)), will be studied below in detail.

**7. Acceleration by the traveling wave**

Let us consider the case when a small ball with the diameter 0.1 μ (the ratio $Z/A = 2.3*10^{-3}$) is accelerated within the range from the initial velocity $\beta = 10^{-3}$ to the final velocity $\beta = 10^{-2}$. The small balls will have such an initial velocity after being pre-accelerated by the high voltage $U = 220$ kV. For further acceleration, two sections, each with an individual feeding RF power and an individual frequency, are required. The velocity in each section will increase by a factor of 3.16, the energy will increase tenfold. In order to slow down the wave in the first section, water with a dielectric permeability $\varepsilon = 80$ will be used, the second section being in vacuum. There is vacuum inside the first section, too. Parameters of the accelerating sections are shown in Table 4.

Table 4. Parameters of the sections.

| Parameter | Section 1 | Section 2 |
|---|---|---|
| $Z/A = 2.3*10^{-3}$ $U_{el.st} = 220$ kV | $P = 6$ MW, $\varepsilon = 80$, $\mu = 1$, partial filling | $P = 1$ MW, $\varepsilon$, $\mu = 1$ without filling |
| Velocity, initial – final, $\beta_{ph}$ | $10^{-3}$ -$3.16*10^{-3}$ | $3.16*10^{-3}$-$10^{-2}$ |
| Starting – ending spiral radius, $r_0$ | 3 – 1.81 cm | 2.41 – 1.39 cm |
| Frequency $f_0$, Hz | $3.333*10^6$ | $12.5*10^6$ |
| Helix pitch distance, h | 0.12 – 0.23 cm | 0.047 – 0.064 cm |
| Average electric field strength, $\bar{E}_0$ | 17 kV/cm | 30 kV/cm |
| Section length* | 1.53 m | 8 m |

* – without the length of the focusing interval

Fig. 1 shows the helix pitch distance (Table 4) as a function of the distance from the beginning of the first section.

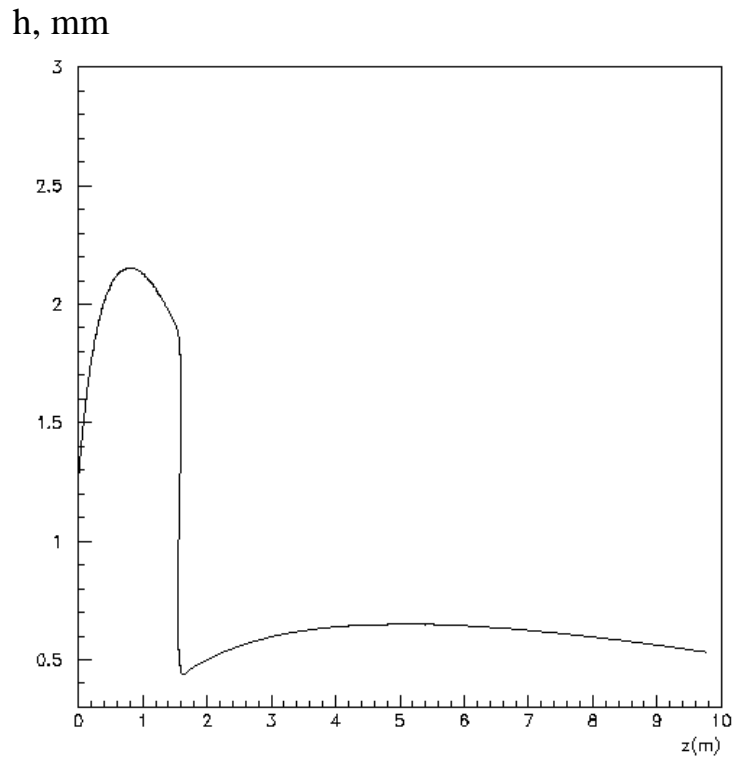

Fig. 1. The helix pitch distance in the first and second sections (Table 4) as a function of the distance from the beginning of the first section. The length of the first section L = 1.53 m.

Fig. 2 shows the dependence of the electric field strength (Table 4) on the distance from the beginning of the first section. Fig. 3 provides the velocity of the small balls in the first and second sections (Table 4) as a function of the distance from the beginning of the first section.

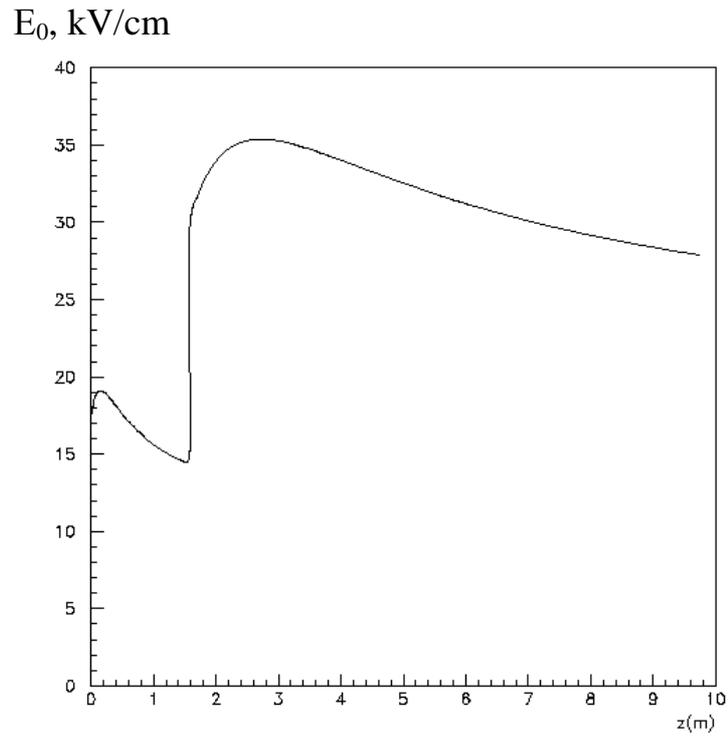

Fig.2. The electric field strength in the first and second sections (Table 4) as a function of the distance from the beginning of the first section.

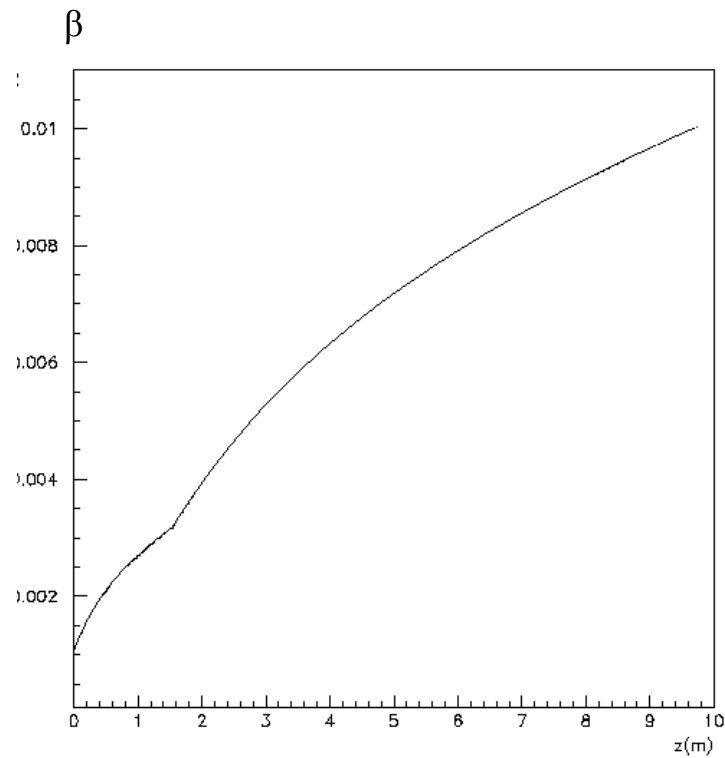

Fig. 3. shows the velocity of the small balls in the first and second sections (Table 4) as a function of the distance from the beginning of the first section. The initial velocity $\beta = 10^{-3}$, the final velocity $\beta = 10^{-2}$.

## 8. Acceleration of particles by the pulse traveling along the spiral structure

A greater value of slowing-down can be achieved using a bigger dielectric permeability $\varepsilon = 1280$, for example, the $TiBaO_3$ ceramic. The parameters of the acceleration section for this case are shown in Table 5. According to formula (14), an RF power of hundreds megawatt is required.

Table 5. Parameters of the sections.

| Parameter | Section 1 | Section 2 |
|---|---|---|
| $Z/A = 2.3*10^{-5}$, dielectric outside the spiral, $U_{el.st.} = 220$ kV | P = 107 MW, $\mu = 1$, $\varepsilon = 1280$ | P = 150 MW, $\mu = 1$, $\varepsilon = 1280$ |
| Velocity, initial – final, $\beta_{ph}$ | $10^{-4} - 3.16*10^{-4}$ | $3.16*10^{-4} - 10^{-3}$ |
| Starting – ending spiral radius, $r_0$ | 10 cm – 5.5 cm | 5.5 cm – 3 cm |
| Frequency $f_0$, Hz | $10^5$ | $6*10^5$ |
| Average electric field strength, $\bar{E}_0$ | 18 kV/cm | 20 kV/cm |
| Helix pitch distance, h | 0.17 – 0.27 cm | 0.27 – 0.4 cm |
| Section length* | 1.5 m | 12.5 m |
| Pulse duration, $\tau$ | 5 µs | 0.83 µs |
| High voltage pulse amplitude, $\tilde{U}_a$ | 170 kV | 220 kV |
| Pulse current amplitude, $\tilde{I}_a$ | 630 A | 680 A |
| Line characteristic impedance $\rho_{wave}$ | 270 Ohm | 323 Ohm |

Fig.4. shows the helix pitch distance (Table 5) as a function of the distance in the first and second sections from the beginning of the first section.

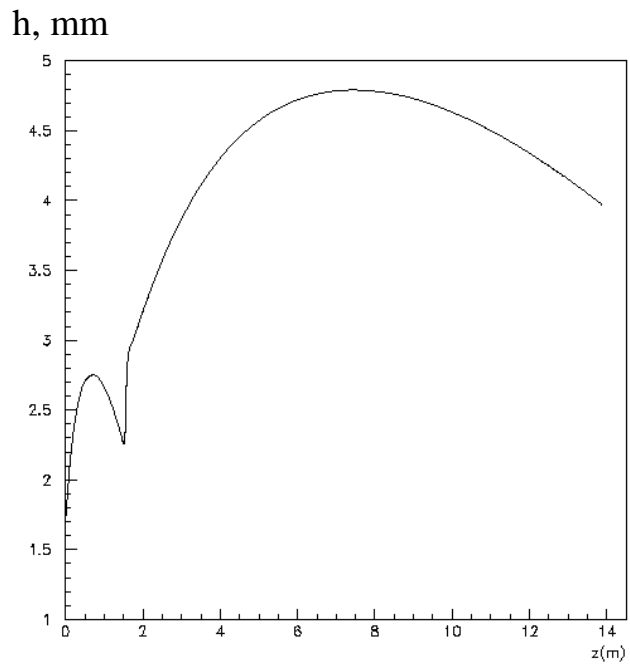

Fig.4. The helix pitch distance (Table 5) as a function of the distance in the first and second sections from the beginning of the first section.

Fig. 5 shows distribution of the electric field strength in the first and second sections as a function of the distance from the beginning the first section.

$E_0$, kV/cm

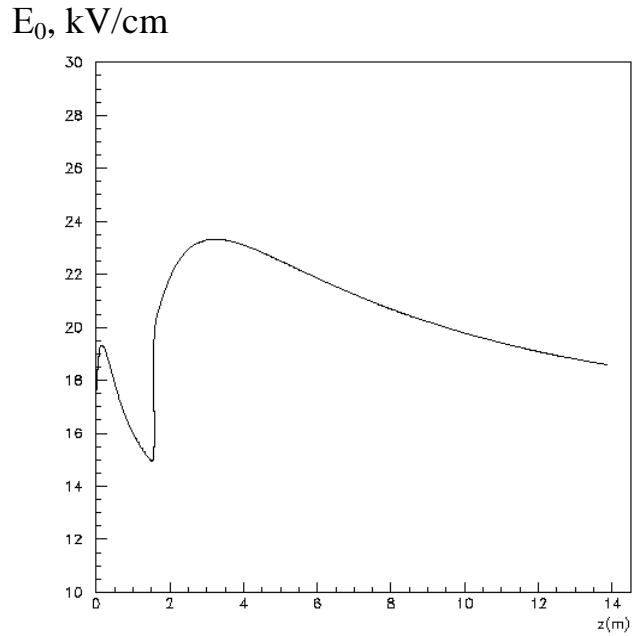

Fig.5. The electric field strength $E_0$ (kV/cm) in the first and second section (Table 5) as a function of the distance z (m) from the beginning of the first section.

Fig.6. provides the velocity in the first and second section (Table 5) as a function of the distance from the beginning of the first section. The initial velocity $\beta = 10^{-4}$, the final velocity $\beta = 10^{-3}$.

$10^2 \beta$

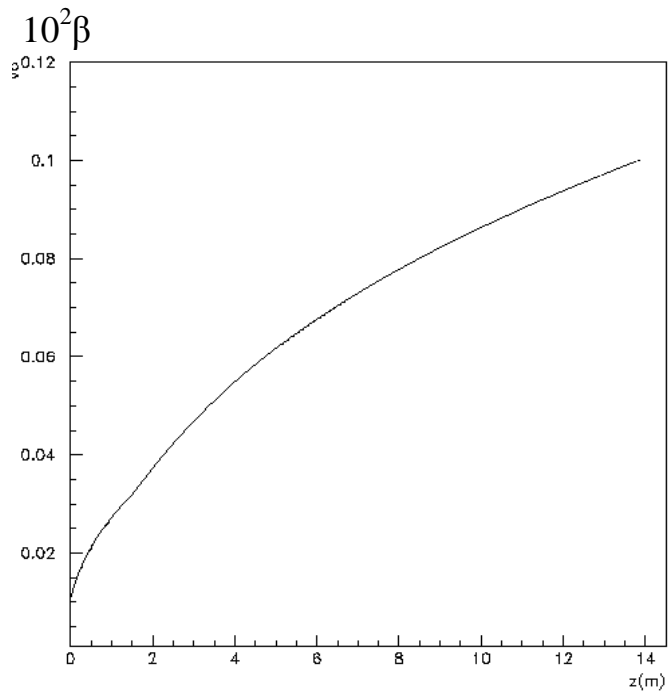

Fig.6. The velocity in the first and second section (Table 5) as a function of the distance from the beginning of the first section. The initial velocity $\beta = 10^{-4}$, the final velocity $\beta = 10^{-3}$.

This power can be easily achieved using the pulse technique. Let us analyze the sine-shaped pulse with the help of the Fourier integral:

$$E = E_0 \sin(2\pi/T_0)t, \qquad (21)$$

where $2\pi/T_0 = \omega_0$, $\omega_0 = 2\pi f_0$. Now we obtain the following:

$$f_1(\omega) = (2/\pi)^{1/2} \int \sin\omega_0 t * \sin\omega t\, dt. \qquad (22)$$

The pulse spectrum is narrow and lies in the frequency range from 0 up to $2\omega_0$. Since practically all waves travel with similar phase velocities in a spiral waveguide, one can expect that the pulse will move without spreading.

Let us introduce the notion of the pulse amplitude $\tilde{U}_a$ connected with the electric field strength in the following way:

$$\tilde{U}_a = E_0 \lambda_{slow}/2\pi, \qquad (23)$$

where $\lambda_{slow} = \beta\lambda$, $\lambda = c/f_0$.

According to the parameters given in Table 5, the electric field strength on the spiral axis $E_0 = 17$ kV/cm, and it is twofold on the spiral surface. Now we can find $\tilde{U}_a = 170$ kV from formula (23). From the Ohm law (equation (14) is a high frequency expression of the Ohm law), one can find:

$$\tilde{I}_a = P/\tilde{U}_a = 630 \text{ A}. \qquad (24)$$

The line characteristic impedance is:

$$\rho_{wave} = \tilde{U}_a/\tilde{I}_a = 270 \text{ Ohm}. \qquad (25)$$

Let us estimate the heating of this spiral line. It is assumed that the spiral is wound up with a tungsten wire of diameter $d_l = 1$mm. In this case, the perimeter of one circle is $2\pi r_0 = 60$ cm, the cross-section of the line is $\pi d_l^2/4 = 7.5*10^{-3}$ cm$^2$, the mass of one circle is m = 9 g, the heat capacity of one gram of tungsten is $C_{tg} = 1347$ J/g*degree, and that of one circle is $C_{circle} = 12$ J/degree.

The resistance of one circle $R_{circle} \approx 3$ Ohm. The current $\tilde{I}_a = 630$ A provides the heating power $W = (1/2)\tilde{I}_a^2 R_{circle} = 0.5*10^6$ W. If the pulse duration $\tau = 5\mu s$, one can obtain $Q_{circle} = W\tau = 2.5$ J. Now, deviding $Q_{circle}$ by $C_{circle}$, one can find that one circle is heated by one pulse by as much as $\Delta T = 0.2$ degrees.

Even larger slowing-down can be achieved by way of decreasing the helix pitch distance according to formula (2). In order to have the behavior of the curves in Fig.4-6 unchanged, it is required to keep the argument of the Bessel functions unaltered: $k_1r_0 \approx k_3r_0 \approx 2\pi r_0/\beta\lambda$, where $\beta$ is the phase and particle velocity, $\lambda = c/f_0$ is the wavelength in the free space.

So, for a tenfold decrease of $\beta = h/2\pi r_0$, one must obtain a tenfold decrease in the frequency $f_0$, and if the spiral radius does not change, a tenfold decrease in the RF power is required, because $(kk_3/k_1^2)$ in formula (14) is simply $\beta$. Parameters of the acceleration sections are given in Table 6.

Table 6. Parameters of the sections.

| Parameter | Section 1 | Section 2 |
|---|---|---|
| $Z/A = 2.3*10^{-7}$, dielectric outside the spiral, $U_{el.st.} = 220$ kV | $P = 10.7$ MW, $\mu = 1$, $\varepsilon = 1280$ | $P = 15$ MW, $\mu = 1$, $\varepsilon = 1280$ |
| Velocity, initial – final, $\beta_{ph}$ | $10^{-5} - 3.16*10^{-5}$ | $3.16*10^{-5}-10^{-4}$ |
| Starting – ending spiral radius, $r_0$ | 10 cm – 5.5 cm | 5.5 cm – 3 cm |
| Frequency $f_0$, Hz | $10^4$ | $6*10^4$ |
| Average electric field strength, $\bar{E}_0$ | 18 kV/cm | 20 kV/cm |
| Helix pitch distance, h | 0.17 – 0.27 mm | 0.27 – 0.4 mm |
| Section length* | 1.5 m | 12.5 m |
| Pulse duration, $\tau$ | 50 µs | 8.3 µs |
| High voltage pulse amplitude, $\tilde{U}_a$ | 170 kV | 220 kV |
| Pulse current amplitude, $\tilde{I}_a$ | 63 A | 68 A |
| Line characteristic impedance $\rho_{wave}$ | 2700 Ohm | 3230 Ohm |

Fig.7 shows the helix pitch distance, Fig.8 provides the electric field strength, and Fig.9 gives the velocity along the first and second sections as a function of the distance from the beginning of the first section.

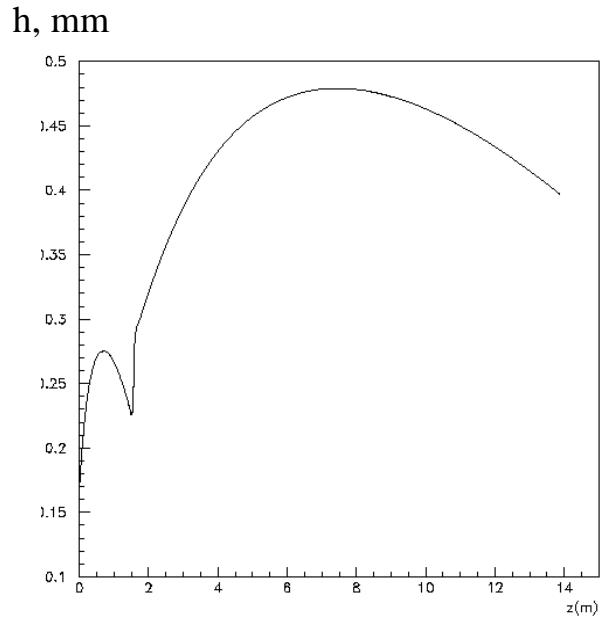

Fig.7. The helix pitch distance in the first and second section (Table 6) as a function of the distance from the beginning of the first section.

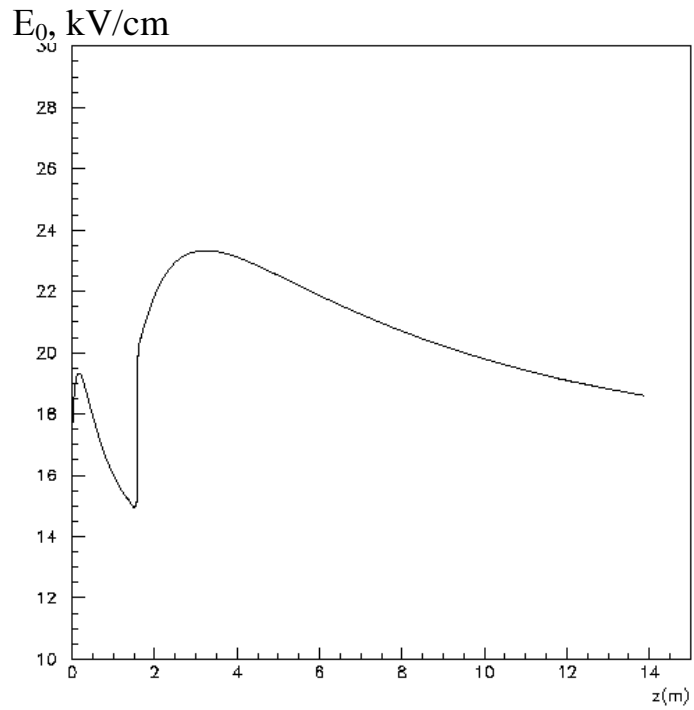

Fig.8. The electric field strength in the first and second section (Table 6) as a function of the distance from the beginning of the first section.

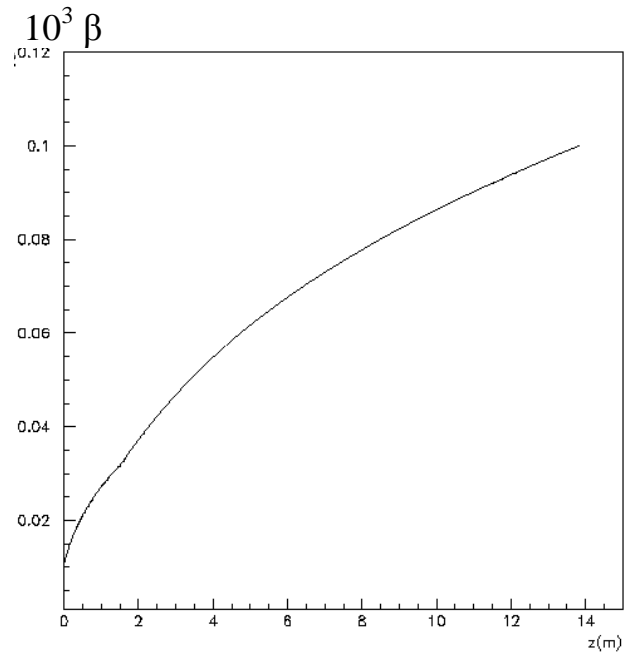

Fig.9. The velocity in the first and second sections (Table 6) as a function of the distance from the beginning of the first section. The initial velocity $\beta = 10^{-5}$, the final velocity $\beta = 10^{-4}$.

In order to increase the helix pitch distance tenfold, with the phase velocity $\beta = 10^{-5}$ preserved constant according to formula (2), the spiral radius must be increased tenfold, and in order to keep all the parameters given in Table 6 and Fig. 6-9 intact, a hundredfold increase in the RF power is required, as indicated by formula (14).

If the helix pitch distance is preserved unchanged for the above increased radius, the phase velocity $\beta = 10^{-6}$ can be achieved. In order to keep the shape of all curves shown in Fig. 6-9 and the electric field strength unaltered, a tenfold increase in the RF power and a hundredfold decrease in frequency, compared to the parameters provided in Table 6, are required. This initial velocity $\beta = 10^{-6}$ can be attained not only by pre-accelerating the small balls with $Z/A = 2.3*10^{-9}$ by passing the high voltage $U_{el.st} = 220$ kV but also using simple centrifugation. The acceleration length for the particles with $Z/A = 2.3*10^{-9}$ will be a hundredfold greater than for the particles with $Z/A = 2.3*10^{-7}$.

## 9. Radial motion

In an azimuthally symmetrical wave, the region of radial defocusing corresponds to the region of phase stability. In this region the radial electric field of the wave accelerates particles in the radial direction, increasing their radial deflections.

Now, let us find the increment of the radial motion. Using the asymptotic of the Bessel function $I_1(x) \approx x/2$ for $x \ll 1$, the equation for the radial motion may be expressed as follows:

$$r'' \approx ZeE_0k_1r/2AM_n, \qquad (26)$$

where: $k_1 = 2\pi/\lambda_{slow}$, $\lambda_{slow} = \beta_z\lambda_0$, $\lambda_0 = c/f_0$, $f_0$ is the frequency of acceleration.

Let us denote the parameter $W_\lambda$ as the ratio of the energy received by a particle from the wave at the wavelength in vacuum to the rest energy of the particle:

$$W_\lambda = ZeE_0\lambda_0/AMc^2. \qquad (27)$$

Then, the frequency is represented by $\omega_r^2 = W_\lambda \pi c^2/\beta_z\lambda_0^2$ and $\omega_0 = 2\pi f_0$, leading to the following result:

$$\omega_r^2 = \omega_0^2(W_\lambda/4\pi\beta_z). \qquad (28)$$

Now, let us calculate $\omega_r^2$ for the parameters:

$Z/A = 2.3*10^{-7}$, $\lambda_0 = c/f_0 = 3*10^6$ cm, $E_0 = 20$ kV/cm, and obtain:

$$W_\lambda = ZeE_0\lambda_0/AMc^2 = 1.38*10^{-5}.$$

As one could expect, this ratio proved to be very small and the increment $\omega_r$ of the radial motion turned out to be of the order of the acceleration frequency value.

The radial deflection of the particles begins to grow exponentially as soon as the particle deviates from the axis. Let us assume that the initial radial deflection of the particle $r_b = 0.3$ cm and the initial radial velocity $V_r = 0$. In fact, the initial radial deflection of the particle can be much smaller but it is taken to be so big only for the purpose of illustrating in detail the radial motion of a particle. The dependence of the particle's radial deflection on the time is:

$$r = r_b e^{\omega_r t}, \qquad (29)$$

and it signifies that the initial radial deflection of the particle will grow rapidly in the absence of outside focusing. The behavior of the particle's motion is linear, depending on the initial deviation. The deviation of a particle with the initial radial deviation ten times smaller is a tenfold-smaller distance. Thus, the focusing element is to be placed through some stretches of the acceleration section.

Let us consider focusing by the electrostatic doublets with the following parameters: the length of the lens $l_l = 7.5$ cm, the distance between the lenses $l_f = 5$ cm, hence, the total length of the entire doublet $l_d = 20$ cm. Let us place the doublet in the first section at the distance $l_{s1} = 30$ cm from its beginning. The second doublet is placed at the distance $l_{s2} = 60$ cm after the first doublet, and the third doublet is placed at the distance $l_{s3} = 60$ cm after the second doublet. So, the total acceleration length in the first section $L_1 = 1.5$ m (column 1, Table 6) is kept the same and three focusing intervals of the total length 0.6 m are added to this section, while the total length of the first acceleration section reaches 1.5+0.6 = 2.1 m, with due account for the length of the focusing interval.

As usual, the motion of two particles located in two orthogonal planes will be further studied. Fig.10 shows the radial deviation of two particles in two orthogonal planes as a function of the distance from the beginning of the first section.

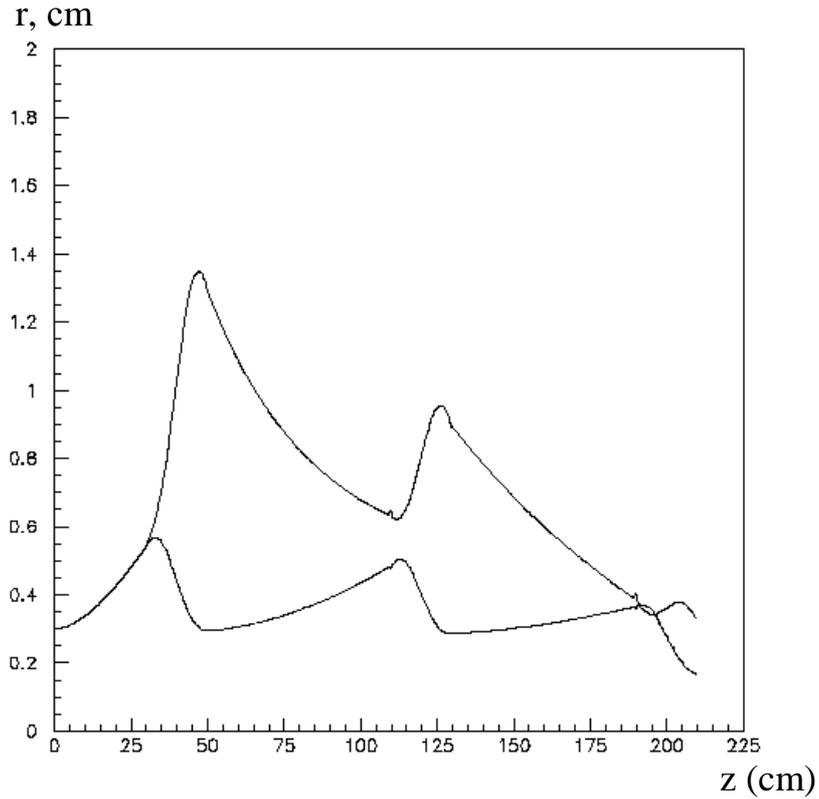

Fig.10. The radial deviation of two particles in two orthogonal planes as a function of the distance from the beginning of the first section.

At the first distance l = 30 cm, the radial deviations of two particles grow exponentially from the initial radial deviation $r_b$ = 0.3 cm up to r = 0.54 cm, according to formula (29). Then, one of the particles increases its radial deviation due to the electric field of the first lens in the doublet, while the other decreases its radial deviation. Upon passing three doublets, the particles will have the same radial deviation as at the beginning, independently of the starting parameters, while their radial deviation has a tendency to decrease. The gradients in the doublets are: $G_{11}$ = 8.3 kV/cm$^2$, $G_{12}$ = - 8.3 kV/cm$^2$, $G_{21}$ = 5 kV/cm$^2$, $G_{22}$ = -5 kV/cm$^2$, $G_{31}$ = 9 kV/cm$^2$, $G_{32}$ = 5.75 kV/cm$^2$.

The second section (Table 6, column 2) is divided into 20 equal intervals, with the distance $l_1$ = 62.5 cm of the acceleration length interchanging with the distance $l_2$ = 20 cm of the focusing length. Focusing of the particles in the second section is conducted using the same doublets with smaller gradients. Fig.11 provides the radial deviation of two particles placed in the orthogonal planes (Table 6) as a function of the distance from the beginning of the first section. The total length of the focusing intervals is 0.2*20 = 4 m and the total length of two acceleration sections increases insignificantly.

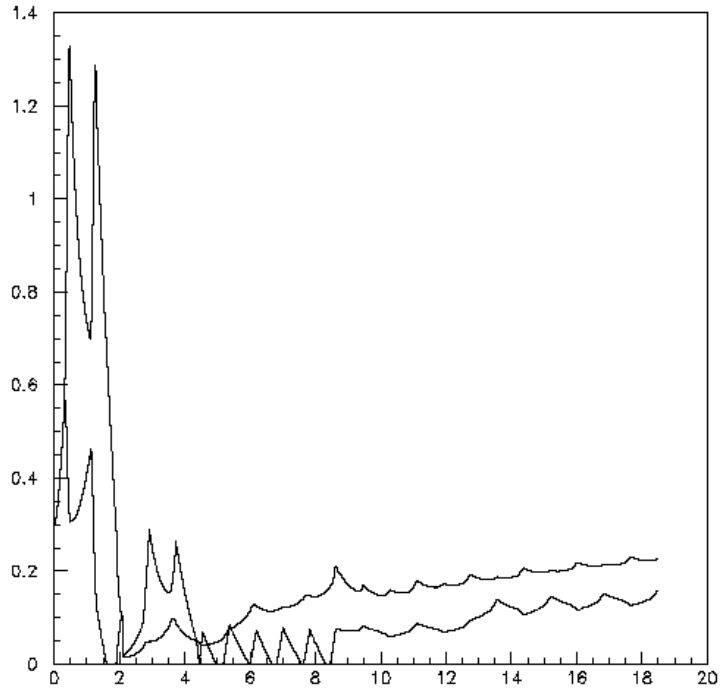

Fig.11. The radial deviation of two particles placed in the orthogonal planes (Table 6) as a function of the distance from the beginning of the first section.

**10. Conclusion**

All correlations obtained in the work have a common principle and can be easily applied to other values of the parameters. For example, a small ball with the ratio $Z/A = 2.3*10^{-8}$ and initial velocity $\beta = 10^{-5}$, $V = 3$ km/s, can have a threefold increase in the velocity at the distance $L = 15$ m (column 1, Table 6).

For these calculations a very big threshold surface voltage $E^i_{th} = 10^{10}$ V/cm has been taken. In case the surface voltage $E_{surf.} = 10^9$ V/cm is achieved, the ratio $Z/A = 2.3*10^{-7}$ will be had by a small ball with the diameter 100 μ, and its motion will be described by the parameters given in Table 6 and curves shown in Fig. 7-11. For $E_{surf.} = 10^8$ V/cm, this ratio will be had by a small ball with the diameter 10 μ. The surface electric field in the region of $E_{surf.} = 10^7$ V/cm typically exists in a multiwire proportional chamber (MWPC).

**References**

1. A.I. Akishin, Cosmic Material Science, Methodic and Tutorial Manual, 2007, Moscow, MSU SRINP, p.154.
2. A.I. Akhiezer, Ya.B. Fainberg, Slow Electromagnetic Waves, UFN, 1951, v. 44, issue 3, p. 321- 367.
3. S.N.Dolya, K.A.Reshetnikova, Heavy Ion Acceleration in a Spiral Waveguide, JINR, R9-2007-120, Dubna, 2007.
4. S.N.Dolya, K.A.Reshetnikova, Two Methods of Injection into the Nuclotron Booster, JINR, R9-2008-120, Dubna, 2008.